\documentclass[english,aps,12pt]{revtex4}
\usepackage[T1]{fontenc}
\usepackage[latin1]{inputenc}
\usepackage{float}
\usepackage{graphicx}
\usepackage{amssymb}

\makeatletter

\usepackage{babel}
\makeatother
\begin{document}

\title{Water confined in nanopores: spontaneous formation of microcavities}

\author{John Russo, Simone Melchionna, Francesco De Luca}

\affiliation{INFM-SOFT, Department of Physics, University of Rome La Sapienza,
P.le A. Moro 2, 00185 Rome, Italy}

\email{simone.melchionna@roma1.infn.it}

\author{Cinzia Casieri}

\affiliation{INFM-SOFT, Department of Physics, University of L'Aquila, Via Vetoio,
67010 L'Aquila, Italy}

\begin{abstract}
Molecular Dynamics simulations of water confined in nanometer sized,
hydrophobic channels show that water forms localized cavities for
pore diameter $\gtrsim2.0\, nm$. The cavities present non-spherical
shape and lay preferentially adjacent to the confining wall inducing
a peculiar form to the liquid exposed surface. The regime of localized
cavitation appears to be correlated with the formation of a vapor
layer, as predicted by the Lum-Chandler-Weeks theory, implying partial
filling of the pore. 
\end{abstract}
\maketitle

\section{Introduction}

Nanotubes \cite{iijima} are important building blocks of nanocomposite
materials and nanomachinery. When immersed in ionic solutions, nanometer-sized
pores can be used for the detection and analysis of electrolytes and
charged polymers, such as DNA \cite{kasianowicz}, and to understand
ion transport and selectivity in biological channels \cite{hille}.
A full understanding of the phase behavior of confined water is a
pre-requisite to interpret adsorption and conductivity data. In particular,
experiments and simulations have shown that, for channels of sub-nanometer
radius water exhibits an intermittent filling/drying transition resulting
from an underlying bi-stable free-energy landscape separated by a
thermally activated barrier, a behavior specific to water and not
observed in a monoatomic fluid \cite{hummer,allen}. The switch between
empty/filled states upon ion translocation has been advocated as a
gating mechanism in passive biological channels \cite{dzubiella,sansom}. 

Cavitation of water under confinement is related to the long-ranged
attractive forces exerted by hydrophobic bodies. Two alternative mechanisms
have been proposed for explaining the occurrence of these forces:
on one hand, the presence of air-filled nanobubbles at the solid surface
drives the attraction via a bubble-driven mechanism; on the other
hand, long-range correlations in the critical region or near the spinodal
line could induce an extended density depletion at the surfaces which
results in the attraction \cite{attard-patey-1,attard-patey-2,israelachvili-pnas}. 

The first line of reasoning relies on the pre-existence of air-filled
cavities at the solid surface \cite{attard}. However, the mechanism
for their stabilization is not fully understood since the Laplace-Young
equation prescribes an internal pressure of the order of $10^{2}$
atm for nanometer-sized bubbles \cite{barrat-hansen}. One question
pertains how such large internal pressure can sustain stable or metastable
cavities, and if these are independently nucleated at the solid, corrugated
surface. It has been suggested that line tension can act to stabilize
such small cavities \cite{pompe}, by reducing the curvature of the
bubble base on the solid surface. As a matter of fact, micron-sized
air bubbles are commonly observed in proximity of hydrophobic corrugated
surfaces and, in the recent literature, there is growing evidence
for the presence of bubbles on corrugated surfaces at the nanoscale.
In particular, recent AFM experiments on silicon oxyde wafer surfaces
of controlled roughness reported on bubble formation at the solid-water
interface \cite{yang}, while syncrotron x-ray reflectivity measurements
reported on a small depletion layer in lieu of localized cavitation
\cite{granick}.

The second interpretation is based on the formation of a vapor layer
in contact with the solid surface. Such depletion layer would involve
a high entropic cost, growing with the extension of the exposed surface,
but compensated by volume-dependent forces arising from molecular
reorganization. The Lum-Chandler-Weeks (LCW) theory \cite{lcw} formulates
the hydrophobic effect in microscopic terms based on the competition
between interfacial and bulk forces which therefore depend on the
solute surface/volume ratio. The molecular mechanism underlying the
force balance is commonly ascribed to the distortion of the hydrogen
bond network. In particular, for small spherical solutes water arranges
in a clathrate structure, while for larger size the ensemble of hydrogen
bond vectors, O-H$^{...}$H, points preferentially towards the solute.
Therefore, if the curvature of the embedded body is low, the distortion
is large and entropic effects prevail over the enthalpic ones, and
vice versa for small solutes. 

The cylindrical pore represents an opposite case to the spherical
solute, where now the confining surface is concave. However, a similar
enthalpic/entropic competition is expected and should depend on the
pore diameter. Thus, a relevant information pertains if, for pore
diameter larger than that characteristic of intermittent behavior,
the liquid phase is completely stabilized or if some peculiar behavior
still takes place. The occurrence of cavitation in water confined
by cylindrical pores has been recently discussed in view of understanding
the hysteresis involved in capillary evaporation \cite{barrat2}. 

We report here the results of Molecular Dynamics (MD) simulations
undertaken on cylindrical, hydrophobic pores of finite length. In
section II the simulation set-up and numerical details are described
and in Section III the Molecular Dynamics data are illustrated. In
Section IV the LCW theory is described and the numerical solutions
are compared with the simulation results. The last Section draws some
conclusions and analyzes recent experimental observations.

\section{Simulation details}

The geometry of the simulation set-up is depicted in Fig. \ref{fig:setup}.
The two pore mouths are connected to the same, periodically folded,
reservoir of quasi-bulk water. The pore diameter $d$ and pore length
$L$ are varied to the following pairs of values $(d,L)=(1.5,4.0)$,
$(2.0,4.0)$, $(2.5,4.0)$, $(2.5,8.0)$, $(3.0,4.0)$ and $(3.5,4.0)\, nm$,
as used to label the runs. Simulations become rapidly costly with
the pore size and we did not attempt to extend our simulations to
larger geometries. Besides the nominal values, the effective cylinder
diameter and length were evaluated via the criterion that the repulsive
wall-oxygen potential be less than $k_{B}T$, giving a reduction of
the effective diameter and increase of effective length by $0.46\, nm$
\cite{allen}. 

The simulation box is periodic in all three directions with dimensions
depending on the pore geometry. The number of water molecules for
all simulations ranges between $852$ and $2464$ units. Water is
represented via the SPC/E computational model \cite{spc}. The confining
oxygen-wall potential is modeled as a smooth surface generated by
carbon atoms distributed uniformly over the surface. The wall-water
interaction acts between each water oxygen and a smooth Lennard-Jones
potential integrated over the dark region of the wall in Fig. \ref{fig:setup}.
Each carbon atom carries a Lennard-Jones potential with parameters
$\sigma=0.345\, nm$ and $\epsilon=0.7294\, kJmol^{-1}$ \cite{allen}
and standard Lorentz-Berthelot rules are used to construct oxygen-wall
interactions. Electrostatic interactions between charged oxygen and
hydrogen atoms are computed with the Ewald method via the Smooth Particle
Mesh Ewald implementation \cite{darden}. Moreover, the confining
medium is taken as non polarizable. 

Special care is taken to keep the reservoir density under control,
such that upon emptying/filling of the channel the reservoir maintains
a bulk-like behavior with an average mass density of $1\, gr/cm^{3}$
away from the pore. This is obtained by varying over time the length
of the reservoir, in direction parallel to the pore axis, by a Berendsen
type of piston \cite{berendsen}, with a characteristic coupling time
of $10\, ps$, during the equilibration and production runs. The feedback
sets the average density in a stripe of thickness $t=10\, nm$ placed
at distance $l=7\, nm$ away from the pore (see Fig. \ref{fig:setup}).
The system temperature is controlled via a Nosè-Hoover thermostat
\cite{frenkelsmit} in order to avoid anomalous temperature drifts
during both the equilibration process and the subsequent production
runs. The system is simulated at $300\, k$ while some auxiliary simulations
are made at $280$ and $320\, K$. The evolution of the system is
followed over times of $100\, ps$ (equilibration) and, subsequently,
of $1\, ns$ (production). Within the simulation time window, stationarity
is monitored by following the number of molecules populating the channel
and the formation and size distribution of cavities. In order to monitor
the effective stationarity of the equilibrated system, the run of
the $(2.5,4.0)\, nm$ geometry has been further extended to $3\, ns$,
without observing any departure from stationarity. 

In order to detect the cavities we have used a coarse-graining procedure
by tesselating the space with cubic cells of edge $0.02\, nm$. An
empty cell is defined by having the distance from any oxygen atom
greater than $0.3\, nm$. The wall position is defined by the largest
radial distance of an oxygen atom from the pore axis and adding an
offset of $0.1\, nm$. A cavity is defined as the cluster of continguous
empty cells which do not belong to the wall region. The distinction
among clusters is made via a graph algorithm and sorted according
to size. In this way, the identity of cavities is clearly established.

\section{Molecular Dynamics results}

The simulation data present two distinct behaviors: for the geometries
$(1.5,4.0)\, nm$ at $T=300\, K$ and $(2.5,4.0)\, nm$ at $T=280\, K$
we observe emptying of the pore, with a large vapor region extending
between the two pore mouths; on the other hand, for the geometries
$(2.0,4.0)$, $(2.5,4.0)$, $(2.5,8.0)$, $(3.0,4.0)$ and $(3.5,4.0)\, nm$,
water persists inside the pore at density close to liquid state, with
formation of nanobubbles in proximity of the confining wall.

At first we report on the process of emptying of the channel, as depicted
in Fig. \ref{fig:longi-empty}. Given an initial, apparently stable
liquid inside the channel, a number of distinct lateral cavities are
formed in proximity of the cylinder wall. Subsequently, one of such
bubbles grows in size, rapidly extending over the whole pore volume.
The final state appears to be stable over the $1\, ns$ time scale.
The homogeneous growth of the vapor region is consistent with the
recently proposed model of capillary evaporation in hydrophobic pores
\cite{barrat2}.

Intermittent emptying/filling of the channel has been previously reported
at $T=300\, K$ for $d\lesssim1.5\, nm$ and $L\lesssim1.0\, nm$
(and eventually for larger diameters at lower temperatures) by using
MD simulations with the same computational model as used here \cite{allen}.
In the previous work, it was found that the time scale of emptying/filling
oscillations is of the order of $\sim100\, ps$. In the current case,
the direct observation of intermittency is prohibitive in terms of
CPU time. In fact, given the larger pore diameters, intermittency
could take place over timescales longer by one order of magnitude,
or more \cite{sansom}. Therefore, we preferred to focus on the interfacial
structuring of the liquid in the filled state.

It is interesting to note that the geometry $(2.4,4.0)\, nm$ exhibits
pore emptying for $T=280\, K$ and pore filling for $T=300\, K$.
The fact that emptying/filling of the pore is sensitive to differences
in temperature as small as $\Delta T=20\, k$ is an indication that
the underlying bulk phase diagram, with the rather narrow region of
bulk liquid water and the close-by liquid-vapor coexistence, is an
important driving factor for the confined fluid. 

For the filled channel, the liquid interface displays the patterns
shown in Figs. \ref{fig:longi-filled} and \ref{fig:trasverse}. The
interface is rather corrugated by the presence of cavities appearing
in proximity of the confining wall (lateral cavities), and rarely
observed as spherical shape located around the cylinder axis. The
cavities resemble spherical caps or rounded cones, with volumes much
larger than the molecular size ($v_{w}\simeq0.027\, nm^{3}$). The
average volumes range between $\langle v\rangle/v_{w}=5.7$ for the
$(d,L)=(2.0,4.0)\, nm$ geometry and $\langle v\rangle/v_{w}=21.0$
for $(d,L)=(3.5,4.0)\, nm$. The peculiar pattern of the liquid exposed
surface is rather representative of the highly cohesive character
of water. It should be noticed that the shape of the vapor cavities
along the transversal section does not appear to be flattened against
the wall, while in the longitudinal direction visual analysis does
not allow to draw a clear conclusion. We infer that line tension effects
can be important only along the longitudinal direction but not on
the transverse direction. In terms of dynamics, the cavities present
high mobility with fast reshaping on the $1\, ps$ time scale. 

The distribution of cavity size is illustrated in the normalized histograms
of Fig. \ref{fig:vPv}. A systematic increase of cavities of larger
volume with the pore diameter is visible, accompanied by a depletion
of smaller, molecular-sized ones. The distribution becomes more shifted
to the right as the diameter of the cylinder grows. In contrast, water
in proximity of a planar hydrophobic surface exhibits cavities with
reduced volumes, basically indistinguishable from fluctuations of
size of the molecular volume $v_{w}$. 

In fig. \ref{fig:vPv_reduced} we report the same histograms but we
filter out the contributions arising from cavities of volume $v<v_{w}$.
By rescaling the abscissa by the cylinder surface area, the histograms
become peaked around the same value $v/(v_{w}\pi dL)\simeq2\, nm^{-2}$,
and width increasing with the pore diameter. However, the integrated
cavity volume per unit surface is mostly contributed by the right
tails of the distributions, which appear to be equal for the $(2.5,4.0)$
and $(3.5,4.0)$ geometries but significantly shifted to the left
for the $(1.5,4.0)$ case. Therefore, cavitation does not seem to
grow further as the diameter is larger than $2.5\, nm$.

The longer channel does not appear to affect the histogram, indicating
that the bubble formation does not depend on interfacial effects arising
from the finite pore length. In principle, the finite length could
affect the liquid/vapor balance and the drying transition. In fact,
by using an elementary macroscopic argument \cite{barrat-hansen},
the difference in free-energy density between the liquid and vapor
states is approximated by $\Delta\Omega/\pi dL=\gamma_{lw}-\gamma_{lv}d/2L$
where $\gamma_{lw}$ and $\gamma_{lv}$ are the liquid-wall and liquid-vapor
surface tensions under ambient conditions, respectively, and bulk
contributions to the free energy have been neglected. In other words,
if finite channel effects are in place, by making the channel longer,
one should move away from the vapor branch. By taking $\gamma_{lw}=7.4\, k_{B}T/nm^{2}$
and $\gamma_{lv}=26\, k_{B}T/nm^{2}$ it is seen that the geometry
$(d,L)=(2.5,4.0)\, nm$ is close to the region where the free-energy
difference changes sign \cite{allen}. Our simulation at room temperature
for $(d,L)=(2.5,8.0)\, nm$ showed that the channel remains filled
as much as the $(2.5,4.0)\, nm$ geometry, without appreciable differences
in the distribution of cavity volumes. Conversely, the sensitivity
to the pore length appears in the fluctuations of the number of adsorbed
water molecules, which approximately drops by a factor two in the
longer channel. We interpret the weak size dependence of cavitation
as due to the corrugated, largely exposed surface of the liquid with
respect to the simple macroscopic model, that renders finite size
effects negligible already at $L=4\, nm$.

\section{Lum-Chandler-Weeks predictions }

In this section we compare our data to the predictions of the LCW
theory for an infinitely long pore. Schematically, the theory can
described as follows. Let us first consider the decomposition of the
microscopic density $n(r)$ into a slowly varying component, $n_{s}(r)$
, and a fast component, $n(r)-n_{s}(r)$. The LCW theory builds upon
the well-known square-gradient theory for liquid-vapor coexistence
\cite{hans} by taking into account self-consistently the fast oscillations
in density. According to the square gradient approximation, the local
grand potential $-W(n)$ is related to the laplacian of the density
via\begin{equation}
\left(\frac{dW}{dn}\right)=m\nabla^{2}n(r)\label{eq:squaregradient}\end{equation}
where $m$ is an effective parameter derived from the underlying interatomic
potential $v(r)$, $m=-\frac{\pi}{6}\int_{0}^{\infty}drr^{4}v(r)$
(Random Phase Approximation \cite{hans}). By applying a coarse graining
procedure, the local density is replaced by the function $n\rightarrow\bar{n}=n+\frac{\lambda^{2}}{2}\nabla^{2}n$
where $\lambda$ is the characteristic length of the coarse graining
procedure. Therefore, the starting equation of the LCW theory is derived
from eq.(\ref{eq:squaregradient}) rewritten in terms of $\bar{n}$,
and asserting that this equations holds for the slow component $n_{s}$,
thus

\begin{equation}
\left(\frac{dW}{dn}\right)_{n_{s}}=\frac{2m}{\lambda^{2}}[\bar{n}-n_{s}]\label{eq:LCW2}\end{equation}
for which we choose $\lambda=0.3\, nm$ and $m=919.9\times10^{21}\, Jnm^{5}mol^{-2}$.
To implement eq. (\ref{eq:LCW2}) one needs prior knowledge of the
bulk equation of state of water near the liquid-vapor coexistence.
Following previous studies \cite{huang}, we used a quartic dependence
plus a linear correction term which models the proximity to coexistence,\begin{equation}
W(n)=\alpha\left(n-n_{l}\right)^{2}\left(n-n_{g}\right)^{2}+\beta(n-n_{l})\label{eq:free-en}\end{equation}
where $\alpha=2.89\times10^{-24}\, Jnm^{9}$, $\beta=3.04\times10^{-29}J$
and the liquid and vapor densities are $n_{l}=32.94\, nm^{-3}$ and
$n_{g}=7.7\times10^{-4}\, nm^{-3}$, respectively \cite{huang}. By
solving the equations in cylindrical coordinates and forbidding any
angular symmetry breaking, we did not attempt to observe the formation
of cavities of given shape. This choice was motivated by the narrow
range of numerical stability found during the solution of the LCW
equations. 

The link between the slowly varying and the complete density fields
is established within the gaussian density fluctuations approximation,
by writing \cite{chandler}\begin{equation}
n({\bf r})=n_{s}({\bf r})-\int d{\bf r}'c({\bf r}')\chi({\bf r},{\bf r}')\label{eq:fredholm}\end{equation}
where the response function $\chi({\bf r},{\bf r}')\equiv\langle\delta n({\bf r})\delta n({\bf r}')\rangle$
is approximated by $\chi({\bf r},{\bf r}')\simeq n_{s}({\bf r})\delta({\bf r}-{\bf r}')+n_{s}({\bf r})n_{s}({\bf r}')h(|{\bf r}-{\bf r}'|)$
and $h(|{\bf r}-{\bf r}'|)-1$ is the bulk pair correlation function.
Moreover, $c({\bf r})$ is the water-wall direct correlation function.
Given $n_{s}({\bf r})$, eq.(\ref{eq:fredholm}) is solved to obtain
$c({\bf r})$ and $n({\bf r})$ via the Nystrom numerical procedure
\cite{press}. The two coupled integro-differential equations (\ref{eq:LCW2})
and (\ref{eq:fredholm}) are solved iteratively. We did not include
the correction to the LCW theory due to the attractive part of the
wall-oxygen potential \cite{huang}. Vice versa, the effective diameter
of MD and the diameter used in LCW calculations, $\bar{d}$, were
considered as equivalent control parameters. 

We have found that for diameter $\bar{d}<2.6\, nm$ the LCW equations
predict complete pore emptying, in agreement with previous simulation
and experimental results, but below our MD data which exhibit emptying
(or eventually intermittency) for effective diameter $\lesssim2.0\, nm$.
However, in this geometry the LCW predictions do not distinguish between
intermittent and cavitating behavior. In the range $2.6<\bar{d}<3.8\, nm$
the theory predicts partial filling of the pore, as illustrated by
the density profiles in Fig. 4, where the micro-phase coexistence
is accompanied by a significant density depletion near the wall. Moreover,
the LCW equations predict a gradual increase of the vapor layer as
the diameter lowers, whilst the MD results show a maximum which varies
slowly with the pore diameter. Following previous authors \cite{huang},
we attribute this difference to the presence of the weak attractive
tail in the water-wall potential. Finally, at larger diameters, the
LCW curves show weakly structured profiles, with a characteristic
non-wetting shape near contact. In essence, the LCW theory predicts
a cross-over between empty and partially filled states for $\bar{d}=2.6\, nm$,
larger than the MD results where the cross-over appears in the $1.5:2.0\, nm$
interval. This discrepancy may be attributed to the one-dimensional
solution of the LCW equations and to the missing attractive tail in
our theoretical treatment, causes which artificially stabilize the
vapor phase, or to the undetermined location of the confining surface,
whose diameter can vary by about $0.5\, nm$.

Notwithstanding the negligible correlations emerging from the LCW
solutions, we can compare MD and LCW data regarding the number of
water molecules filling a channel of length $4\, nm$ as a function
of the channel effective diameter $\bar{d}$. From Fig. \ref{fig:Water-count}
it is apparent that the LCW model shows systematically lower values
than the MD data in the region where the theory exhibits partial filling.
The MD data are intermediate between the LCW values and the simple
model of uniform filling of the pore at the water liquid density.
As for the discrepancy in the cross-over diameter, the lack of longitudinal
symmetry breaking and the weak attractive tail in the simulated water-wall
potential, might explain the smaller number of water molecules predicted
by the LCW theory.

\section{Conclusions}

The behavior of water in contact with an extended hydrophobic surface
has attracted considerable attention on account of the intrinsic thermodynamic
problem, which has significant implications for protein folding \cite{huang},
for ionic transport in biological channels \cite{hille} and for the
boundary conditions for fluid flow in microchannels \cite{degennes}.

Our study clearly demonstrates the spontaneous formation of vapor
microcavities on the nanometer scale for water in contact with a concave
cylindrical surface of diameter $d\gtrsim2.0\, nm$. The cavities
are localized both in the angular and, more importantly, in the longitudinal
directions. To our knowledge, this surprising result is the first
observation for water confined in such geometry. 

The results are particularly interesting for the on-going debate on
the hydrophobic effect, in which a number of different interpretations
have been put forward, such as entropic effects due to molecular rearrangement,
electrostatic effects and spontaneous cavitation due to the metastability
of the fluid \cite{israelachvili-pnas}. Recent experiments have focused
on analyzing water in contact with a hydrophobic surface and found
contrasting results. In particular, either cavitation \cite{yang,ishida}
or an extended low-density depletion layer \cite{granick} have been
observed. The fact that water around a concave, but smooth, hydrophobic
surface forms short-ranged islands of vapor, modulated by the surface
curvature, suggests that the contrasting experimental observations
arise from the sensitive structural response of water to the roughness
of the solid surface. 

Moreover, our study sheds some light on the experimental observation
of hysteresis in water intrusion/extrusion cycles in pores and the
interpretation based on homogenous nucleation \cite{barrat2}. The
simulation data underline the metastable character of highly confined
water. However, the rich structural patterns exhibited by the interface
are absent in a simple fluid and, therefore, are unlikely to be explained
in terms of a macroscopic approach. When comparing the simulation
data with a quasi-microscopic treatment, namely the Lum-Chandler-Weeks
theory, the density profiles and the cross-over diameter between empty
and partially filled states of the latter were found to agree only
qualitatively with the simulation data. The discrepancies were explained
on the basis of the one-dimensional solution of the LCW equations. 

Finally, we wish to comment on some recent measurements on ionic conductance
in nanopores of diameter $d\simeq10\, nm$ showing that transport
displays an anomalous response, with a five orders of magnitude reduction
in the current spectral density power and a strongly noisy response
with respect to bulk behavior \cite{dekker}. This has been attributed
to the presence of spherically shaped air bubbles trapped inside the
nanopores such that the translocating ions encounter a two-phase filled
region. Although the differences in pore diameter between the presently
simulated and the experimental systems might seem large, one might
expect that water cavitation is still effective in the wider pores
or that cavitation is locally enhanced by the translocating ion. The
effect of such cavitation would induce non trivial wall-ion dielectric
interactions, modulated by the imperfect screening of water, and non
trivial hydrodynamic forces. Such scenario would imply a coupled ion-bubble
transport mediated by the microscopic liquid-vapor coexistence.

\section{Acknowledgements}

We wish to thank Jean-Pierre Hansen and Francesco Sciortino for critical
reading of the manuscript and Mauro Chinappi for help with the simulation
setup.

\pagebreak

\section*{Figure Captions}

\begin{figure}[H]

\caption{\label{fig:setup}Schematic representation of a cut through the simulation
cell. The dark area is the confining cylindrical wall and the shaded
area is the region where water is maintained at mass density of $1\, gr/cm^{3}$
(see text for details).}
\end{figure}

\begin{figure}[H]

\caption{\label{fig:longi-empty}Snapshots of the $(1.5,4.0)\, nm$ channel
at four different instants separated by $10\, ps$ showing the process
of pore emptying. The pictures are produced with the surface generation
algorithm of the VMD software \cite{vmd} by taking into account the
oxygen atoms only. }
\end{figure}

\begin{figure}[H]

\caption{\label{fig:longi-filled}Snapshot of an instantaneous configuration
of the $(2.5,4.0)\, nm$ channel, generated as for Fig.\ref{fig:longi-empty}. }
\end{figure}

\begin{figure}[H]

\caption{\label{fig:trasverse}Evolution of cavities (shaded regions) formed
in the $(2.5,4.0)\, nm$ pore at $300\, K$, and analyzed on a slab
of thickness $0.1\, nm$ placed at the midpoint of the pore axis.
The four snapshots (a,b,c,d) correspond to successive configurations
separated by $1\, ps$. The white region corresponds to the wall region,
grey to water filled regions, black to the empty regions (see text
for details).}
\end{figure}

\begin{figure}[H]

\caption{\label{fig:vPv}Frequency of occurrence of cavities with volume $v$
normalized by the volume per water molecule ($v_{w}=0.027\, nm^{3}$)
for $(d,L)$ equal to $(2.0,4.0)\, nm$ (solid line), $(2.5,4.0)\, nm$
(dashed line), $(2.5,8.0)\, nm$ (long dashed line), $(3.5,4.0)\, nm$
(dot-dashed line) and for water close to an infinite plane (dotted
line). All data are at $T=300\, K$. To highlight the dependence on
the pore geometry, the normalized frequency of occurrence $P(v)$
is multiplied by the volume. The inset displays the unscaled frequency
(same symbols of the main plot). }
\end{figure}

\begin{figure}[H]

\caption{\label{fig:vPv_reduced}Histograms as in Fig. \ref{fig:vPv}, but
for cavity volumes larger than $v_{w}$ . The three curves refer to
$(d,L)=(2.0,4.0)$, $(2.5,4.0)$ and $(3.5,4.0)\, nm$, with line
styles as in Fig. \ref{fig:vPv}. Abscissa have been divided by the
cylinder surface area. }
\end{figure}

\begin{figure}[H]

\caption{Radial density profiles divided by the density of liquid water $n_{l}$.
The radial coordinate is rescaled with the effective radius (MD data)
and the value $\tilde{d}/2$ (LCW data). Upper panel: MD profiles
at $300\, k$ for nominal pore diameter $d=2.0\, nm$ (black), $2.5\, nm$
(red), $3.0\, nm$ (green) and $3.5\, nm$ (blue) and length $L=4\, nm$.
Middle panel: LCW profiles for $\tilde{d}$ equal to $2.8\, nm$ (black),
$3.0\, nm$ (red), $3.4\, nm$ (green), $3.6\, nm$ (blue). Lowner
panel: LCW profiles for $\tilde{d}$ equal to $3.8\, nm$ (black),
$4.0\, nm$ (red), $6.0\, nm$ (green), $10.0\, nm$ (blue).}
\end{figure}

\begin{figure}[H]

\caption{\label{fig:Water-count}Number of water molecules occupying channels
of length $4\, nm$ obtained from MD (filled circles), LCW solutions
(open squares) and a uniform filling of the channel (dashes curve).
The vertical jump in the LCW curve signals the cross-over between
empty and partially filled states. }
\end{figure}

\pagebreak

\pagebreak

\begin{center}
\includegraphics[clip,angle=90,width=0.7\columnwidth,keepaspectratio]{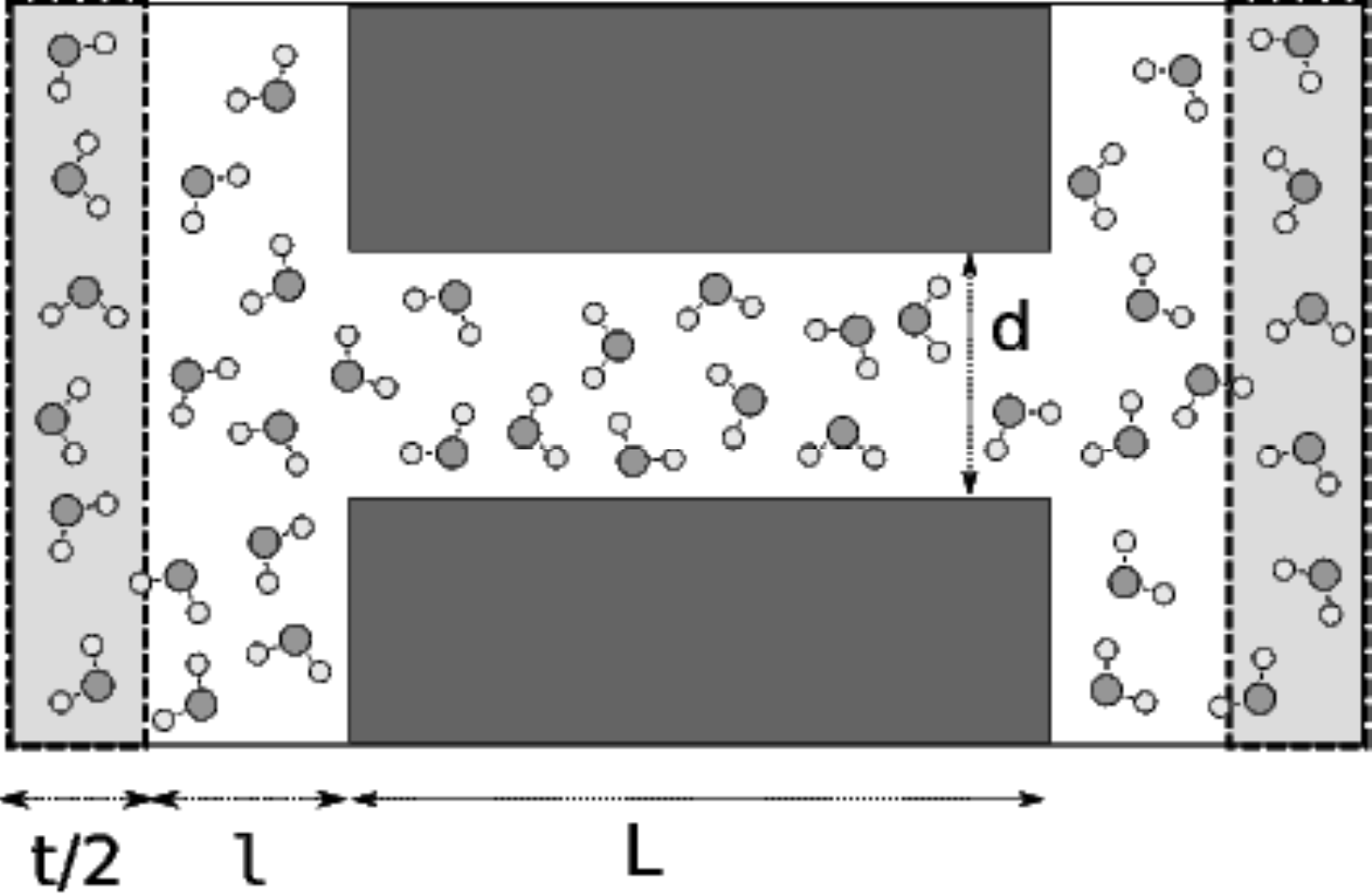}
\par\end{center}

\begin{center}
Figure 1, Russo et al.
\par\end{center}

\pagebreak

\begin{center}
\includegraphics[clip,width=0.7\columnwidth,keepaspectratio]{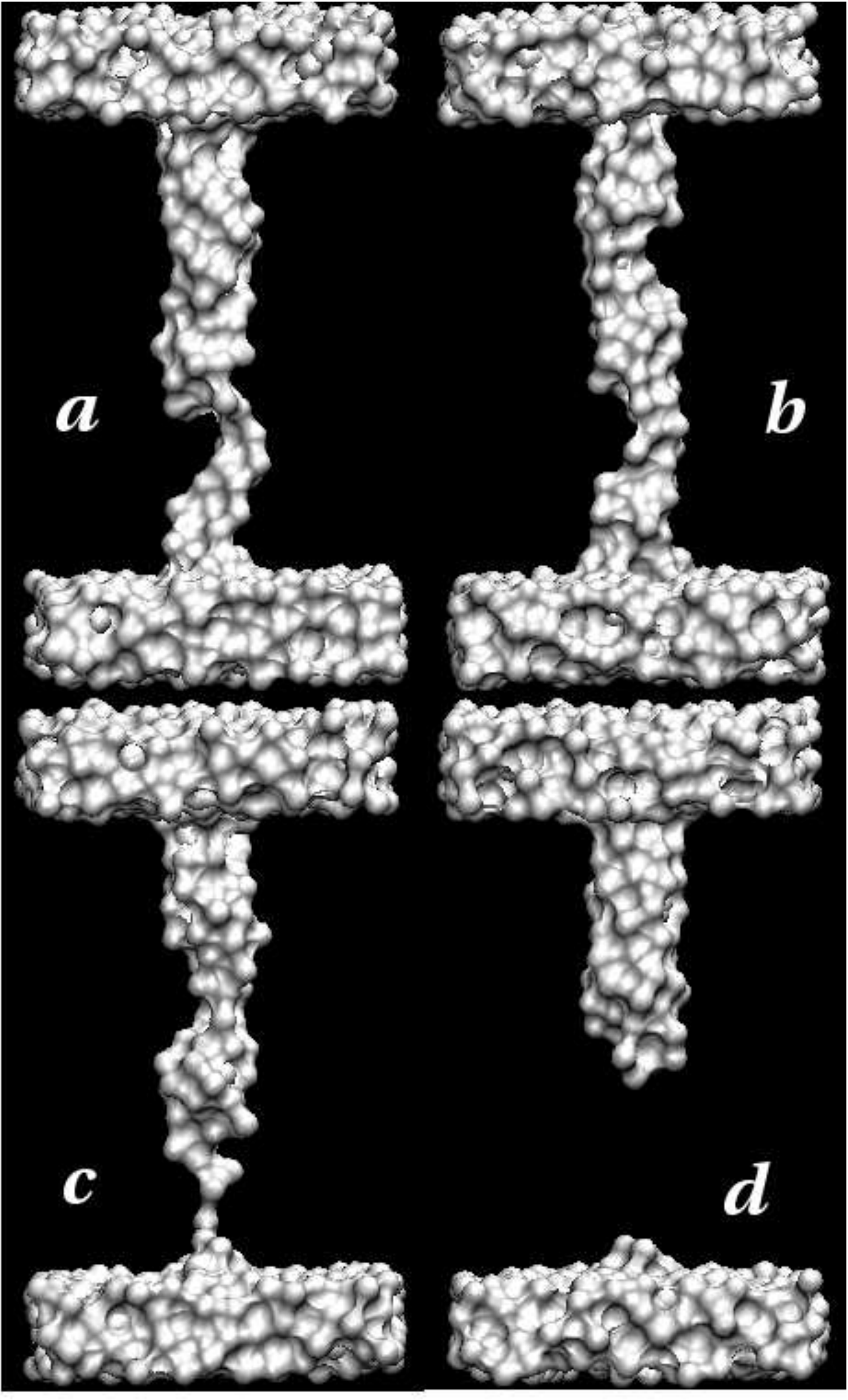}
\par\end{center}

\begin{center}
Figure 2, Russo et al.
\par\end{center}

\pagebreak

\begin{center}
\includegraphics[clip,width=0.7\columnwidth,keepaspectratio]{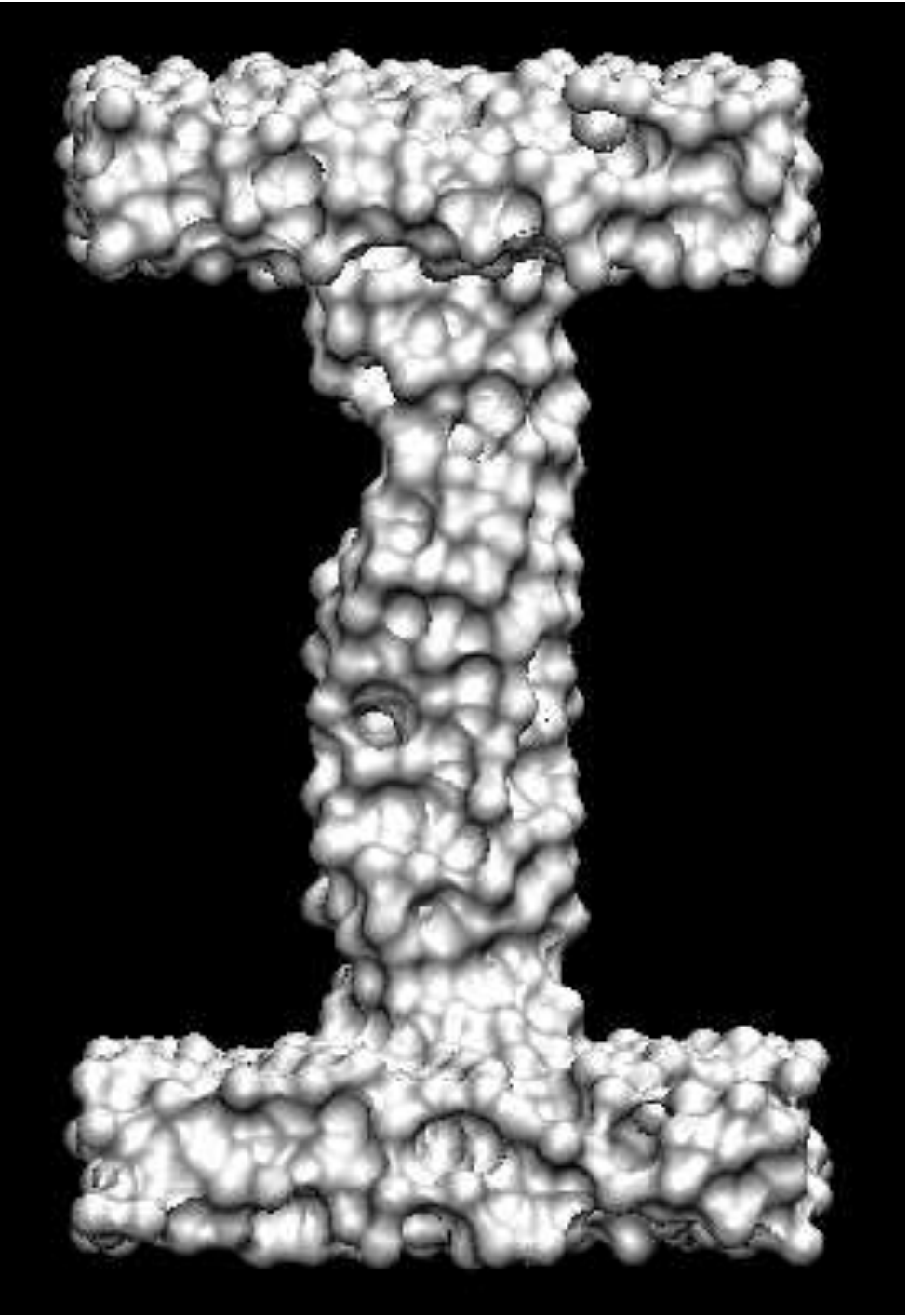}
\par\end{center}

\begin{center}
Figure 3, Russo et al.
\par\end{center}

\pagebreak

\begin{center}
\includegraphics[clip,width=0.8\columnwidth,keepaspectratio]{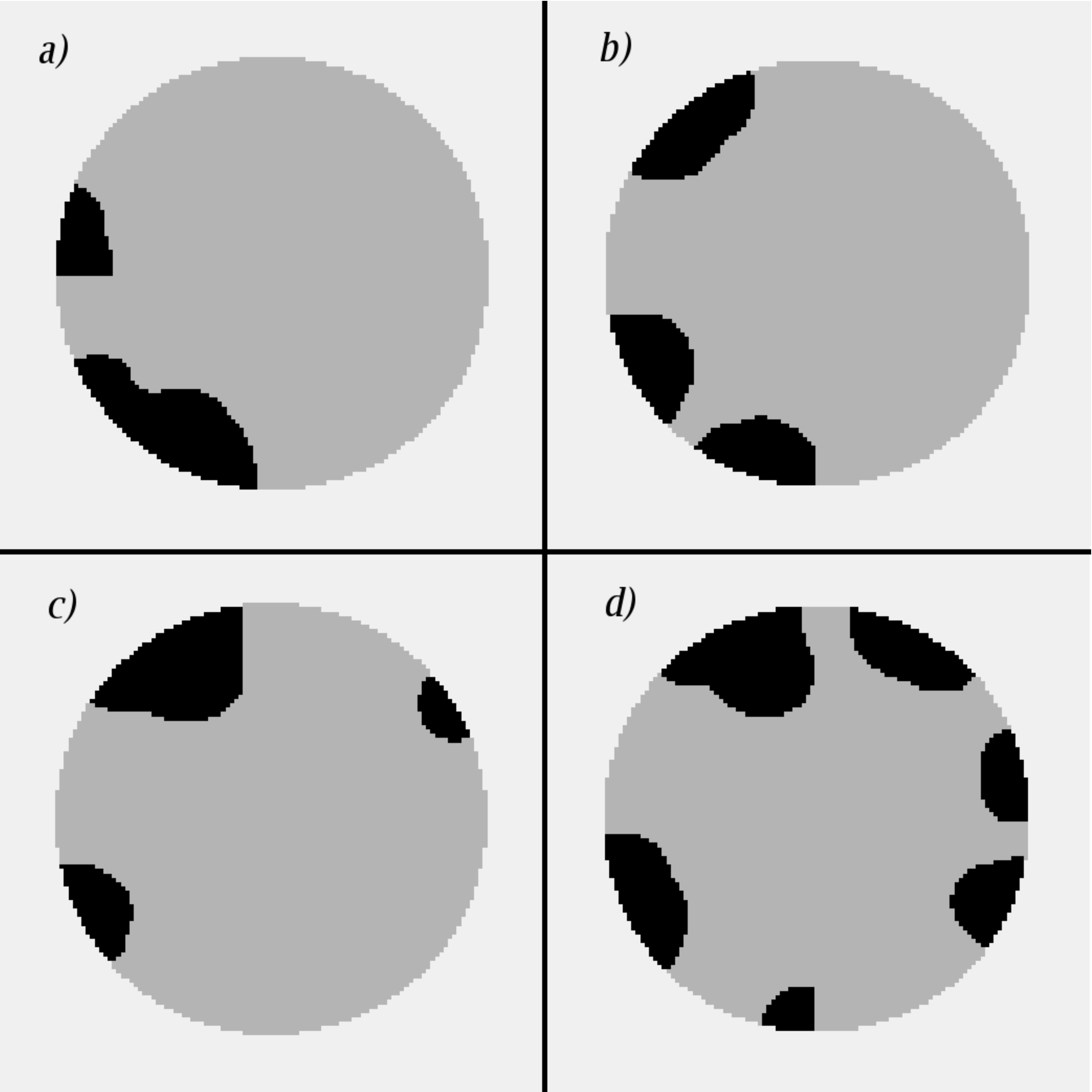}
\par\end{center}

\begin{center}
Figure 4, Russo et al.
\par\end{center}

\pagebreak

\begin{center}
\includegraphics[clip,width=0.8\columnwidth,keepaspectratio]{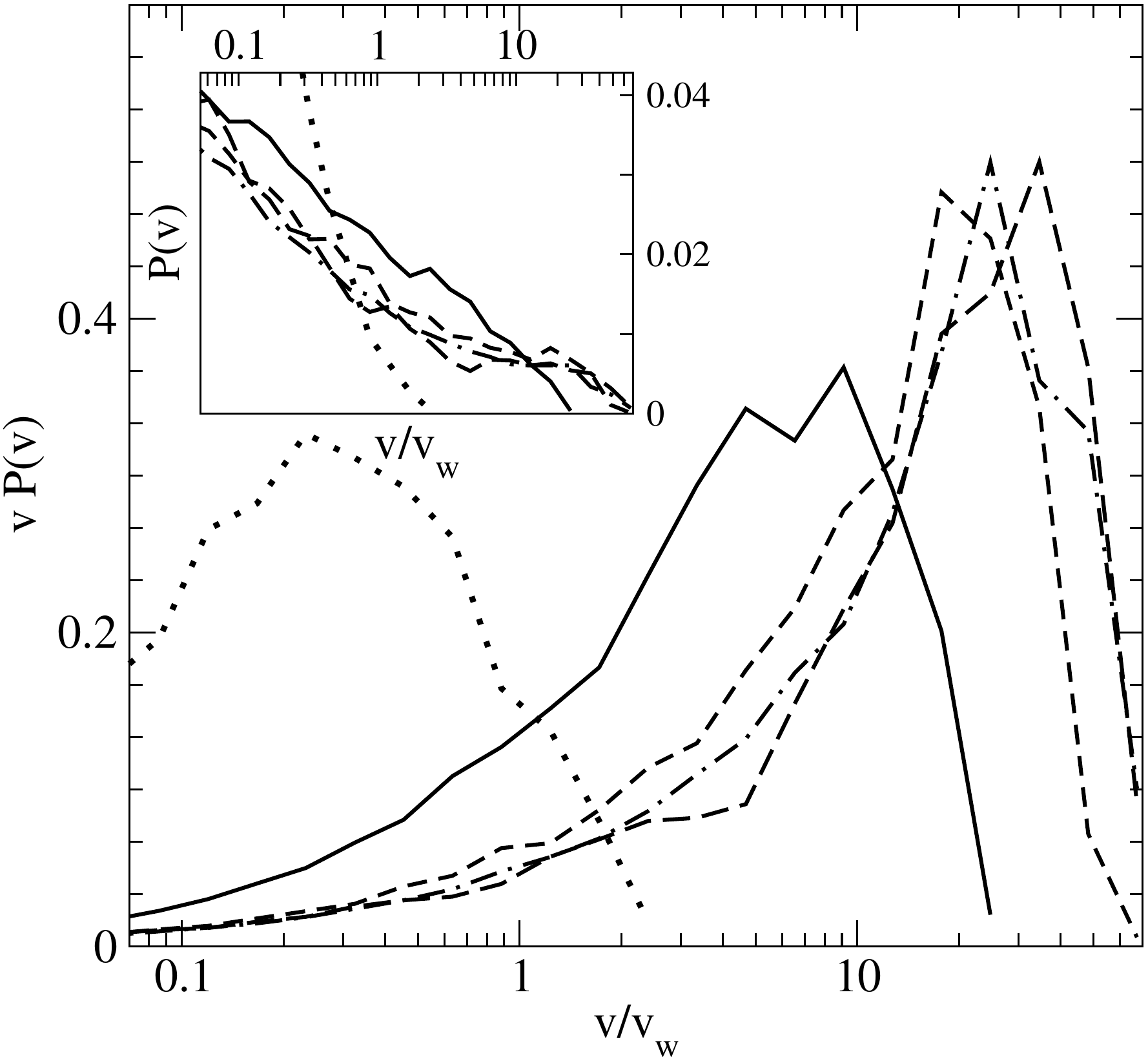}
\par\end{center}

\begin{center}
Figure 5, Russo et al.
\par\end{center}

\pagebreak

\begin{center}
\includegraphics[clip,width=0.8\columnwidth,keepaspectratio]{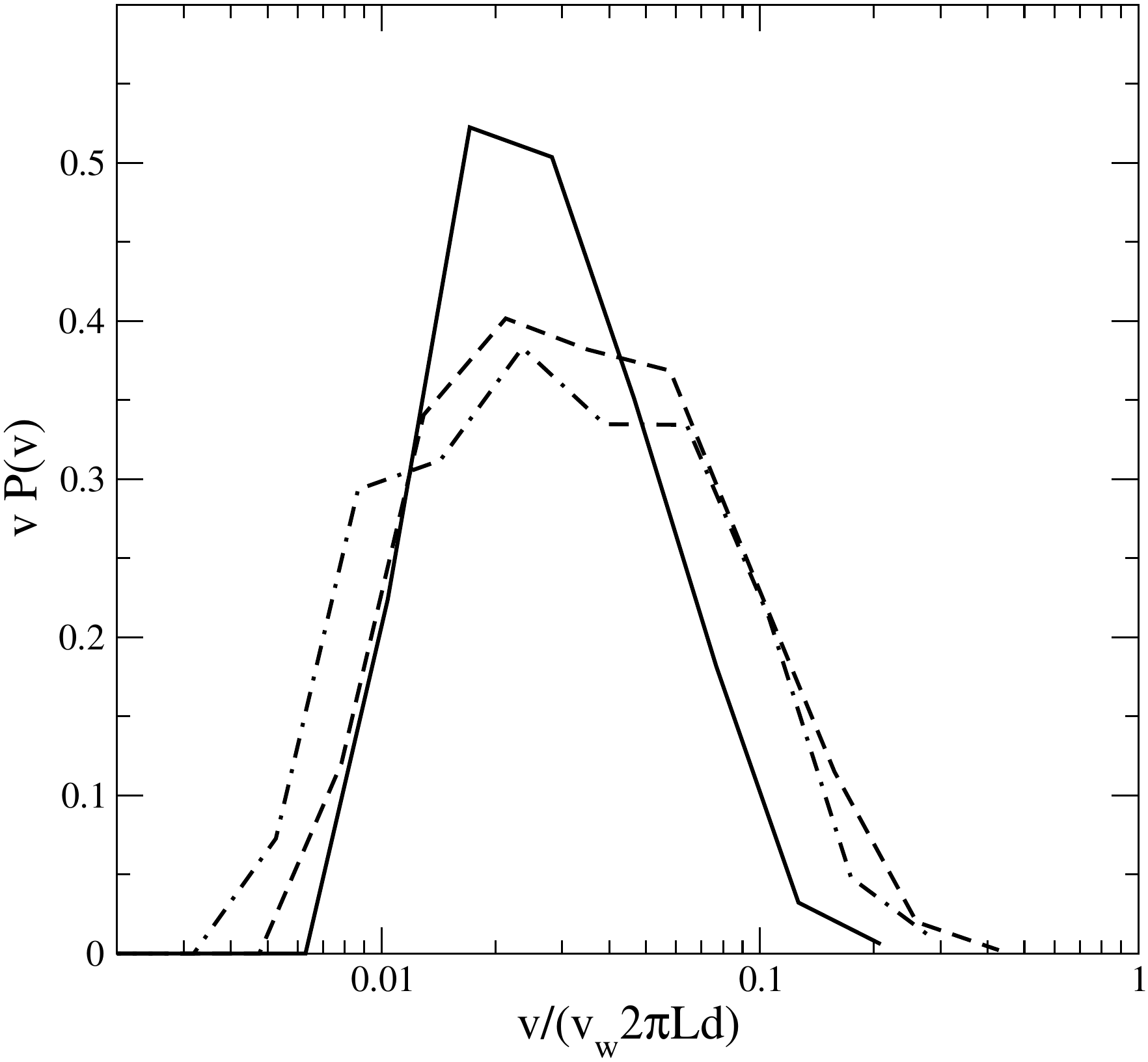}
\par\end{center}

\begin{center}
Figure 6, Russo et al.
\par\end{center}

\pagebreak

\begin{center}
\includegraphics[clip,width=0.9\columnwidth]{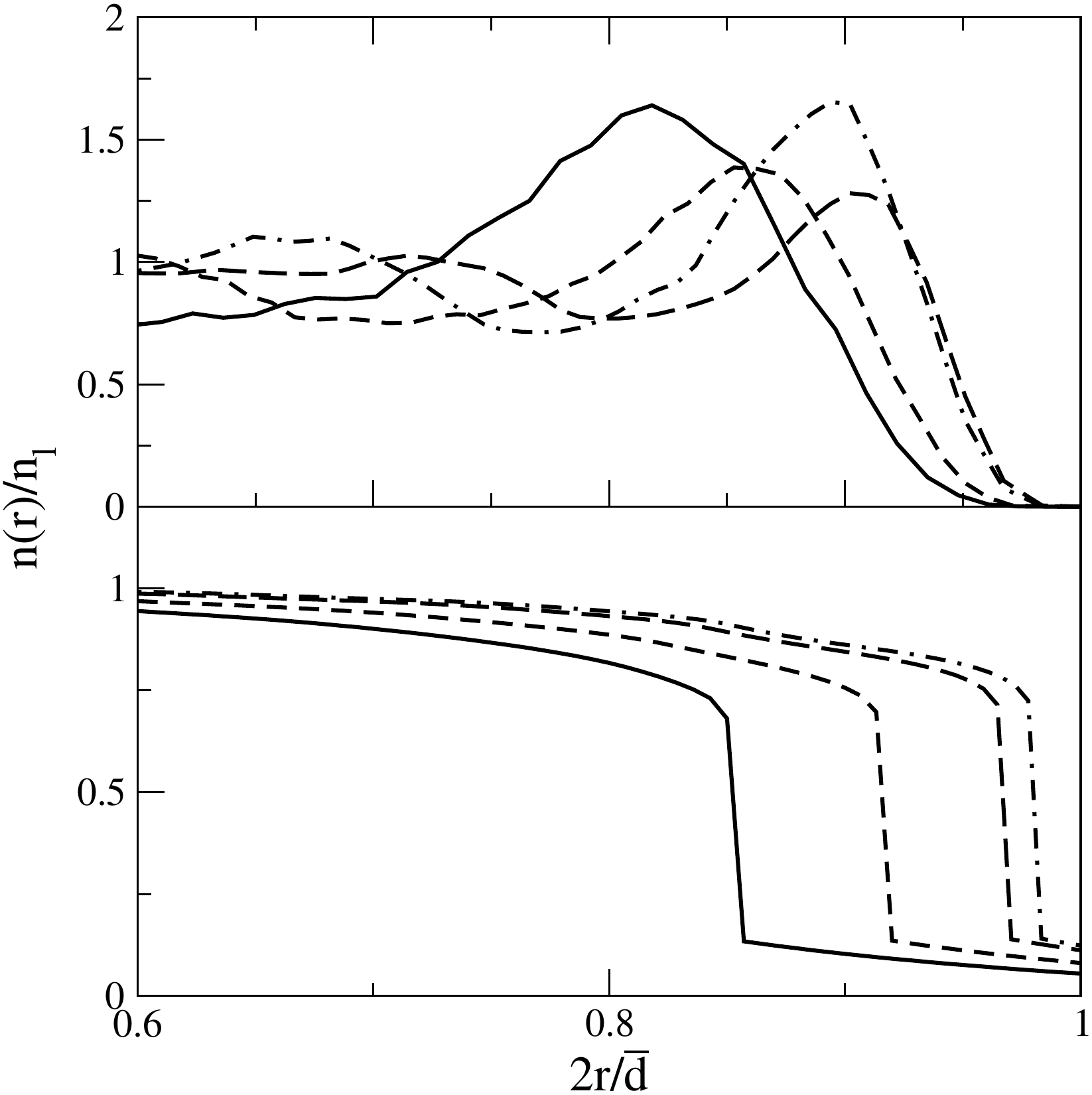}
\par\end{center}

\begin{center}
Figure 7, Russo et al.
\par\end{center}

\pagebreak

\begin{center}
\includegraphics[clip,angle=90,width=0.9\columnwidth]{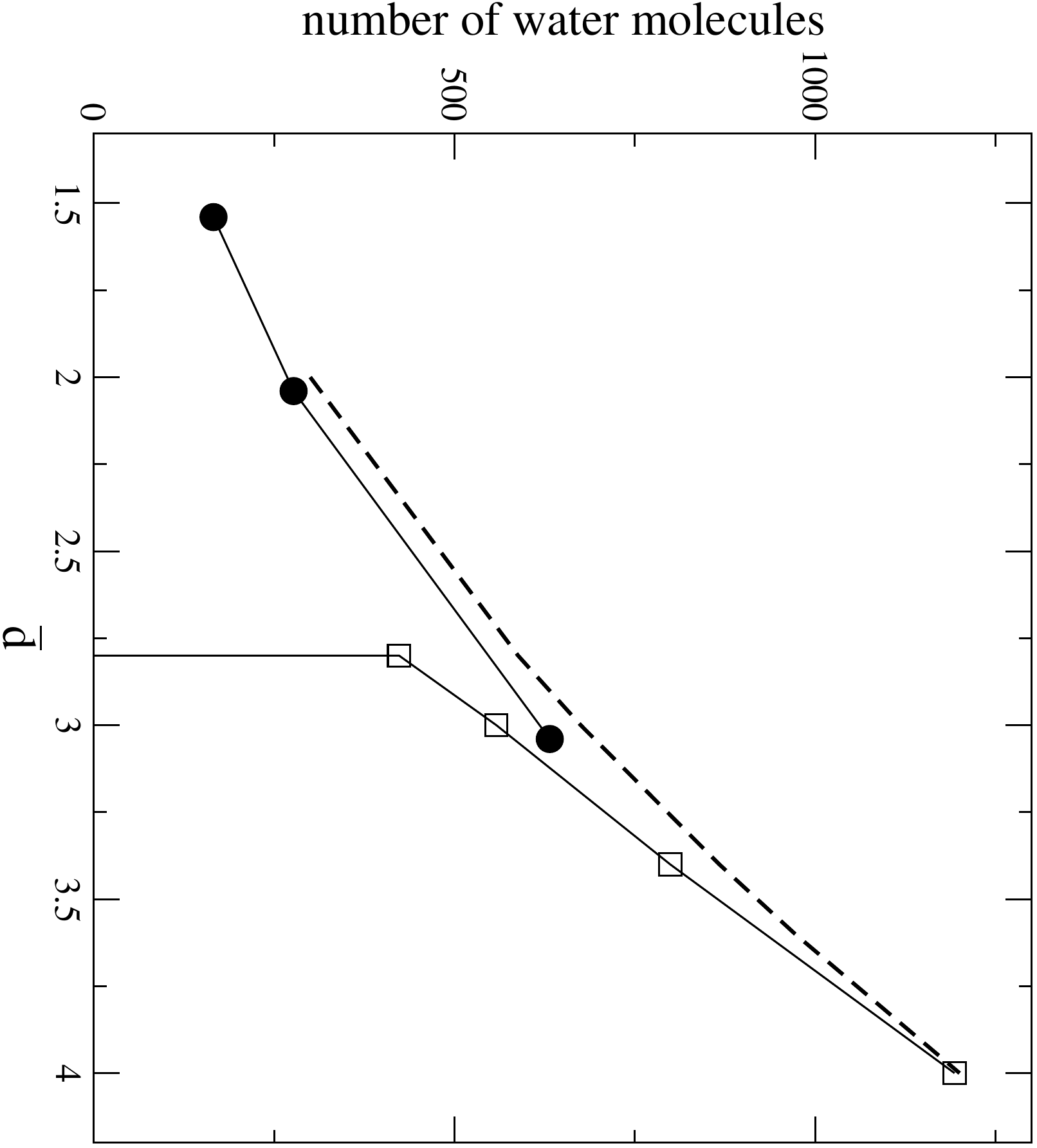}
\par\end{center}

\begin{center}
Figure 8, Russo et al.
\par\end{center}
\end{document}